\documentclass[preprint2,numberedappendix]{emulateapj-rtx4}
\usepackage{bm}
\newcommand{\G}{\,{\rm G}}
\newcommand{\cm}{\,{\rm cm}}
\newcommand{\Mm}{\,{\rm Mm}}
\newcommand{\Eq}[1]{Equation~(\ref{#1})}
\usepackage{color}

\begin{document}
\preprint{NORDITA-2015-25}

\title{Evolution of magnetic helicity and energy spectra of solar active regions}

\author{Hongqi Zhang$^1$, Axel Brandenburg$^{2,3,4,5}$ and D.D. Sokoloff$^{6,7}$}
\affil{
$^1$Key Laboratory of Solar Activity, National Astronomical Observatories, Chinese
Academy of Sciences, Beijing 100012, China, \\
$^2$Nordita, KTH Royal Institute of Technology and Stockholm University,
Roslagstullsbacken 23, SE-10691 Stockholm, Sweden,\\
$^3$Department of Astronomy, AlbaNova University Center,
Stockholm University, SE-10691 Stockholm, Sweden\\
$^4$JILA and Department of Astrophysical and Planetary Sciences,
University of Colorado, Boulder, CO 80303, USA\\
$^5$Laboratory for Atmospheric and Space Physics,
University of Colorado, Boulder, CO 80303, USA\\
$^6$Department of Physics, Moscow University, 119992 Moscow, Russia, \\
$^7$Pushkov Institute of Terrestrial Magnetism, Ionosphere and Radiowave Propagation \\
of the Russian Academy of Sciences, Troitsk, Moscow, 142190, Russia
}

\submitted{\today,~ $ $Revision: 1.100 $ $}

\begin{abstract}
We adopt an isotropic representation of the Fourier-transformed two-point
correlation tensor of the magnetic field to estimate the magnetic energy
and helicity spectra as well as current helicity spectra of two individual
active regions (NOAA~11158 and NOAA~11515) and the change of the spectral
indices during their development as well as during the solar cycle.
The departure of the spectral indices of magnetic energy and current
helicity from $5/3$ are analyzed, and it is found that it is lower than
the spectral index of the magnetic energy spectrum.
Furthermore, the fractional magnetic helicity tends to increase when
the scale of the energy-carrying magnetic structures increases.
The magnetic helicity of NOAA~11515 violates the expected hemispheric sign rule,
which is interpreted as an effect of enhanced field strengths at scales
larger than $30$--$60\Mm$ with opposite signs of helicity.
This is consistent with the general cycle dependence, which shows that
around the solar maximum the magnetic energy and helicity spectra are steeper,
emphasizing the large-scale field.
\end{abstract}

\email{hzhang@bao.ac.cn}

\section{Introduction}

Many aspects of the continuous regeneration of the
global (large-scale) magnetic
field of the Sun can be explained by a helical turbulent dynamo,
as originally suggested by \cite{Par55} and \cite{SKR66}.
Attempts at finding evidence for a helical magnetic field in the Sun
go back to \cite{Seehafer90}, who found that the
force-free alpha parameter as a proxy of the current
helicity of the Sun is predominantly negative in the northern hemisphere
and predominantly positive and the southern.
This has been confirmed in many subsequent studies
\citep{Pevtsov94,Ab97,Bao98,Chae01,Hagino04,zetal10}.

The magnetic field generation by a large-scale dynamo process is
expected to have opposite signs at large and small length scales
\citep{KR82,Zeldovich83,See96,Ji99}.
Such a field is called bihelical.
Observational evidence for a bihelical field has been obtained by
in situ measurements in the solar wind \citep{BSBG11}, but the
sign of the helicities at the large- and small-scale components
turned out to have opposite signs to what was expected.
This change of sign was confirmed in subsequent theoretical work by
\cite{WBM11,WBM12}, but it is unclear how close to the solar surface
this change of sign occurs.

Here by large-scale fields we mean those parts that characterize
the global or large-scale dynamo process in the Sun.
The remaining parts are automatically referred to as small-scale fields.
In mean-field dynamo theory \citep{Par55,SKR66,Moffatt78}, the global
field of the Sun can be described by azimuthal averaging, which immediately
implies that most of the fields in active regions vanish under such
averaging and would therefore constitute part of the small-scale field.
In practice, however, one would like to replace azimuthal averaging
by some kind of Fourier or spectral filtering.
The relevant filtering scale is not known a priori, but it would roughly
coincide with the scale at which the magnetic helicity changes sign.
The results of the present paper suggest that this scale might be
$30$--$60\Mm$, corresponding to wavenumbers in the range
$k=2\pi/(30$--$60\Mm)=0.1$--$0.2\Mm^{-1}$.
In this sense, small scales refer to the scale of active regions and,
of course, all smaller scales down to the dissipation scale \citep{St12}.

The technique used to obtain the scale dependence of magnetic helicity
through observations goes back to \cite{MGS82}, who made the assumption of
isotropy to express the Fourier transform of the two-point correlation
tensor of the magnetic field in terms of magnetic energy and helicity
spectra.
Their approach made use of one-dimensional spectra obtained from
timeseries of measurements of all three magnetic field components.
The Taylor hypothesis was therefore used to relate the two-point
correlation function in time to one in space \citep{Taylor38}.
In the work of \cite{Zhang2014}, again the assumption of isotropy
was made, but a full two-dimensional array of magnetic field vectors
was used, so no Taylor hypothesis was invoked.
\cite{Zhang2014} applied this technique to the active region NOAA~11158
to determine magnetic energy and helicity spectra.
The current helicity spectrum has been estimated from the magnetic
helicity spectrum, again under the assumption of isotropy, and
its modulus shows a $k^{-5/3}$ spectrum at intermediate wavenumbers.
A similar power law is also obtained for the magnetic energy spectrum.
Both were found to be consistent with turbulence simulations \citep{BS05b}.

The variation of magnetic energy and helicity spectra
of active regions with the solar cycle is another important aspect.
Observational evidence for changes of the integrated current helicity of
active regions with the solar cycle has been studied before
\citep[cf.][]{zetal10,YangSB12,Zhang13,PP14}, but changes of their spectral
properties with the solar cycle still remain an open question.

In the present paper we consider the evolution of the spectrum
of magnetic helicity and its relationship with the magnetic energy from
photospheric vector magnetograms of two solar active regions,
NOAA~11158 and NOAA~11515.
We also present the change of statistical properties of magnetic
energy and helicity spectra, the mean magnetic helicity and energy
densities, as well as the typical length scales of active
regions over the solar cycle.

\section{Basic formalism}

As explained in \cite{Zhang2014}, we introduce the two-point correlation
tensor of the magnetic field $\bm{B}({\bm x},t)$,
$\langle B_i({\bm x},t) B_j({\bm x}+{\bm\xi},t)\rangle$, and
write its Fourier transform with respect to ${\bm\xi}$ as
\begin{equation}
\left\langle\tilde{B}_i({\bm k},t)\tilde{B}_j^*\!({\bm k}',t)\right\rangle
=\Gamma_{ij}({\bm k},t)\delta^2({\bm k}-{\bm k}'),
\end{equation}
where the tildes indicate Fourier transformation, i.e.,
$\hat{B}_i({\bm k},t)=\int B_i({\bm x},t)\,e^{i{\bm k}\cdot{\bm\xi}}d^2\xi$
and the asterisk denotes complex conjugation.
Under isotropic conditions, the spectral correlation tensor
$\Gamma_{ij}({\bm k},t)$ takes the form \citep{Zhang2014}
\begin{equation}
\Gamma_{ij}(\bm{k},t)=\frac{2E_M(k,t)}{4\pi k}(\delta_{ij}-\hat{k}_i\hat{k}_j)
+\frac{iH_M(k,t)}{4\pi k}\varepsilon_{ijk}k_k,\label{eq:helispec5}
\end{equation}
where $E_M(k,t)$ and $H_M(k,t)$ are the magnetic energy and helicity spectra,
respectively, hats denote unit vectors, and $k=(k_x^2+k_y^2)^{1/2}$
is the wavenumber.
We have ignored the permeability factor in the definition of $E_M(k,t)$,
which is here measured in units of $\G^2\Mm=10^8\G^2\cm$.
Note that the expression for $\Gamma_{ij}({\bm k},t)$ differs from
that of \cite{Moffatt78} by a factor $2k$, because in two dimensions
the differential for the integration over shells in wavenumber space
changes from $4\pi k^2\,dk$ to $2\pi k\,dk$.

As shown in \cite{Zhang2014}, $E_M(k,t)$ and $H_M(k,t)$ are obtained as
\begin{eqnarray}
2E_M(k,t)&=&2\pi k\,\mbox{Re}\left\langle \Gamma_{xx}+\Gamma_{yy}+\Gamma_{zz}\right\rangle_{\phi_k},
\label{EMdef} \\
kH_M(k,t)&=&4\pi k\,\mbox{Im}\left\langle\cos\phi_k\Gamma_{yz}-\sin\phi_k\Gamma_{xz}\right\rangle_{\phi_k},
\end{eqnarray}
where the angle brackets with subscript $\phi_k$
denote averaging over annuli in wavenumber space.

Of particular importance in turbulence theory is the scale of the
energy-carrying eddies or in this case the energy-carrying magnetic
structures.
It is defined as a weighted average of the inverse wavenumber over
the magnetic energy spectrum, i.e.,
\begin{eqnarray}
\label{eq:integspec} 
l_M=\left.\int k^{-1}E_M(k)\,dk\right/\int E_M(k)\,dk.
\end{eqnarray}
and is in turbulence theory commonly referred to as the integral scale.
The realizability condition $k|H_M(k,t)|\le2E_M(k,t)$ \citep[cf.][]{Moffatt69}
can be rewritten in integrated form \citep[e.g.][]{TKBK12} as
\begin{eqnarray}
|{\cal H}_M(t)|\le 2l_M{\cal E}_M(t),
\end{eqnarray}
where ${\cal E}_M(t)=\int_0^\infty E_M(k,t)\,dk$ is the mean
magnetic energy density and ${\cal H}_M(t)=\int_0^\infty H_M(k,t)\,dk$
is the mean magnetic helicity density.
The integrated realizability condition allows to then define the
fractional (nondimensional) magnetic helicity density as
\begin{eqnarray}
r_M={\cal H}_M/2l_M{\cal E}_M,\label{eq:integspecr}
\label{rM_and_lM}
\end{eqnarray}
which varies between $-1$ and $1$ and is therefore
sometimes also referred to as the relative helicity. It must not be
confused with the gauge-invariant magnetic helicity of \cite{Berger84},
which is sometimes also called relative magnetic helicity.
Because of the conservation of magnetic helicity \citep{Woltjer58},
if the turbulence is left to decay, $|r_M|$ will tend to unity
\citep{Taylor86} after a time that depends on its initial value
as was demonstrated in simulations \citep{TKBK12}.

\section{Magnetic Helicity and Energy Spectra of Individual Active Regions}
\subsection{Active Region NOAA~11158}
 
\begin{figure}
\begin{center}
\includegraphics[width=0.5\textwidth]{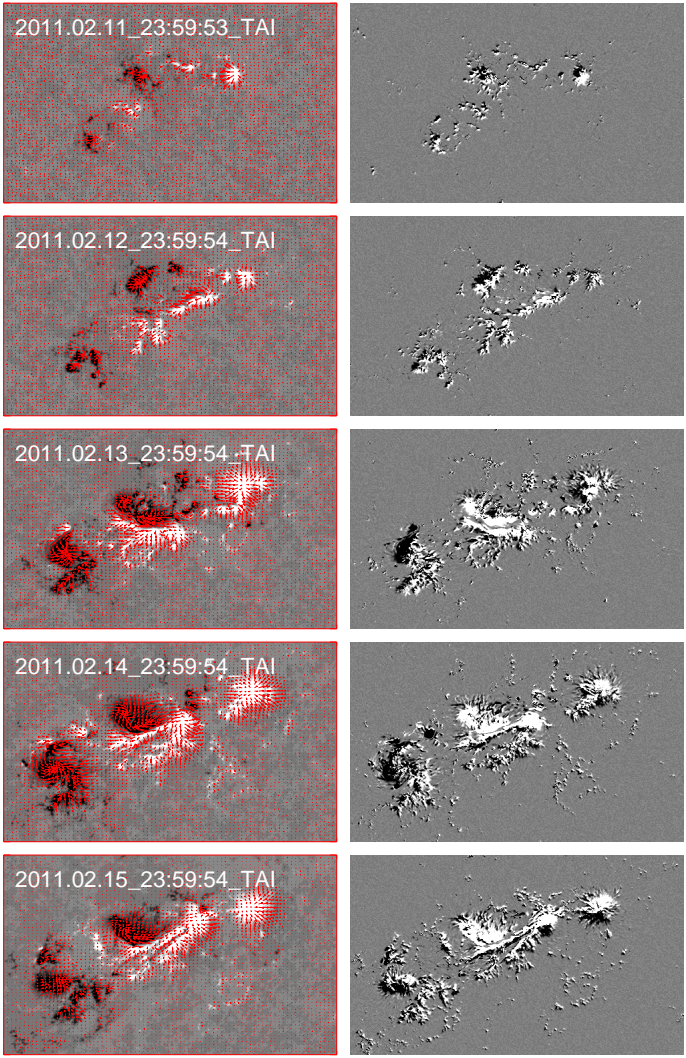}
\end{center}
\caption{
Photospheric vector magnetograms (left) and maps of $J_zB_z$ (right)
of the active region NOAA~11158 between 2011 February 11 to 15.
The arrows indicate the transverse component of the magnetic field.
Light (dark) shades indicate positive (negative) values of $B_z$ on the
left and $J_zB_z$ on the right.
}\label{fig:helispec1}
\end{figure}

We have analyzed data from the solar active region NOAA~11158 from
2011 February 12 to 16 at approximately $13\degr$ southern latitude,
which was taken by the Helioseismic and Magnetic Imager (HMI)
on board the {\em Solar Dynamics Observatory} ({\em SDO}). The pixel resolution
of the magnetograms is about $0''\!\!.5$, and the field of view is $250''\times
150''$ in Figure \ref{fig:helispec1}.
In our study 600 vector magnetograms of the active region have been used.

Figure \ref{fig:helispec1} shows the photospheric vector magnetograms and
the corresponding distribution of $h_C^{(z)}=J_z B_z$ from the vector
magnetograms of that active region on different days. Here $J_z=\partial
B_y/\partial x-\partial B_x/\partial y$ is proportional to the vertical
component of the current density. The superscript `$(z)$' on $h_C^{(z)}$
indicates that only the vertical contribution to the current helicity
density are available.

\begin{figure}
\begin{center}\hspace*{-10mm} 
\includegraphics[width=0.45\textwidth]{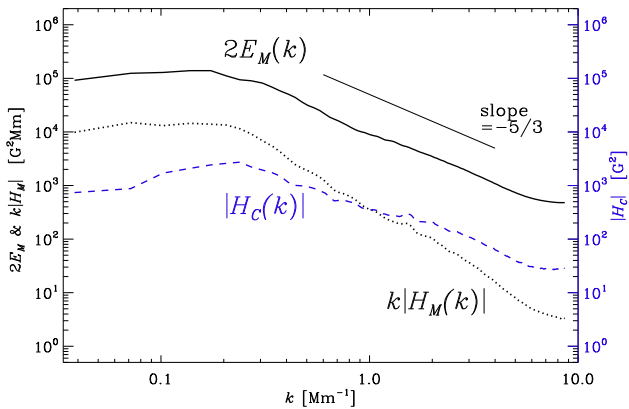}
\end{center}
\caption{
$2E_M(k)$ (solid line), $k|H_M(k)|$ (dotted line), and
$|H_C(k)|$ (dashed line) obtained by averaging over 600 vector magnetograms of
active region NOAA~11158 from 2011 February 12 to 16.
}\label{fig:helispectot1}
\end{figure}

We now consider the magnetic energy and helicity spectra for a field of view
of $256''\times 256''$ (i.e., $512\times 512$ pixels).
We average the resulting spectra from 600 vector magnetograms
from 2011 February 12 to 16; see Figure \ref{fig:helispectot1}.
These are comparable with that of \cite{Zhang2014} except that the
fluctuations in the calculation of individual samples are now reduced
by the averaging.
This provides a basic estimate of the spectra of magnetic
energy and helicity of this active region.

\begin{figure}
\begin{center}
\includegraphics[width=0.5\textwidth]{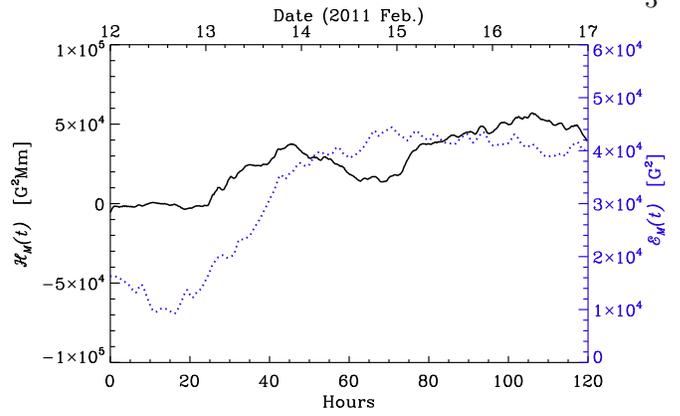}
\end{center}
\caption{
Evolution of mean photospheric magnetic helicity density
${\cal H}_M(t)$ (solid line) and mean magnetic energy density
${\cal E}_M(t)$ (dotted line) of active region NOAA~11158.
}\label{fig:helispec3}
\end{figure}

Figure~\ref{fig:helispec3} shows the evolution of the mean magnetic
helicity and energy densities for NOAA~11158,
obtained by integrating over all $k$.
It is found that the mean magnetic helicity and energy densities first
increase and then continue to change as the active region develops.
An intermediate decrease of the mean magnetic helicity density
of the active region occurred
on 2011 Feb.~14, while the mean magnetic energy density did not decrease.
This shows that the mean magnetic helicity density in the active region does not
always have a monotonous relationship with the mean magnetic
energy density.
This is also consistent with the trends found by \cite{GZZ12} for
the current and kinetic helicity densities for NOAA~11158
using vector magnetograms and subsurface velocity fields.

According to the theory of hydromagnetic turbulence by \cite{GS95}, the
magnetic energy spectrum has a power-law inertial range of $\propto k^{-\alpha}$,
where the spectral index $\alpha$ is compatible with 5/3 (about 1.67)
and is dominated by contributions from wave vectors in the direction
that is perpendicular to the local magnetic field.
These spectral properties were confirmed by \cite{Ab05} and \cite{St12}
based on solar magnetic field observations.

\begin{figure}
\begin{center}
\hspace*{-15mm}
\includegraphics[width=0.45\textwidth]{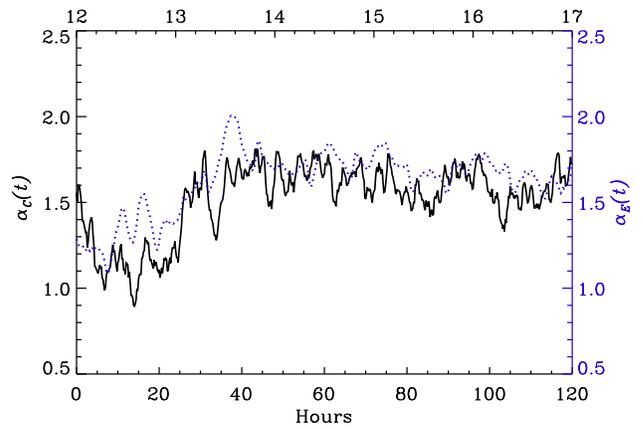}
\end{center}
\caption{
Evolution of $\alpha_C$ (solid line) and $\alpha_E$ (dotted line) of
active region NOAA~11158.
}\label{fig:helispec4}
\end{figure}

We estimate the spectral indices $\alpha_i$ with $i=E$ for magnetic energy
and $i=C$ for current helicity within the wavenumber interval
$1\,{\rm Mm}^{-1}<k<6\,{\rm Mm}^{-1}$, which should capture the spectral
behavior in the inertial range of the turbulent magnetic field
(we postpone the discussion of different wavenumber intervals to
Section~\ref{ResolutionDep}).
Figure~\ref{fig:helispec4} shows the evolution of $\alpha_i$ for NOAA~11158.
It is found that the minimum $\alpha_E$ is 1.1 and the maximum is 2.0.
After 2011 February 13 as the active region became more developed,
the mean value was about 1.65.

Under isotropic conditions, $H_C(k,t)$ is related to $H_M(k,t)$ via
$H_C(k,t)\approx k^2H_M(k,t)$.
Figure~\ref{fig:helispec4} shows the evolution of $\alpha_C$ for NOAA~11158.
It is found that the minimum $\alpha_C$ is 0.9 and the maximum is 1.8.
After 2011 February 13, the mean value was about 1.7.
This implies that the value of $\alpha_C$ of this active region is of
the order of 5/3 and consistent with our previous study \citep{Zhang2014}.
Furthermore, the mean values of $\alpha_E$ and $\alpha_C$
of this active region at the solar surface are roughly consistent
with a $k^{-5/3}$ power law.
We also consider the spectrum $kH_M(k)$ and the
corresponding spectral index $\alpha_{kH}$.
If we had perfect power-law scaling, then $\alpha_{kH}=\alpha_C+1$,
but the actual fits tend to give slightly larger values:
its minimum value is 1.9, the maximum is 2.8,
and the mean value is 2.65 as the active region develops.

\begin{figure}
\begin{center}\hspace*{-10mm}
\includegraphics[width=0.45\textwidth]{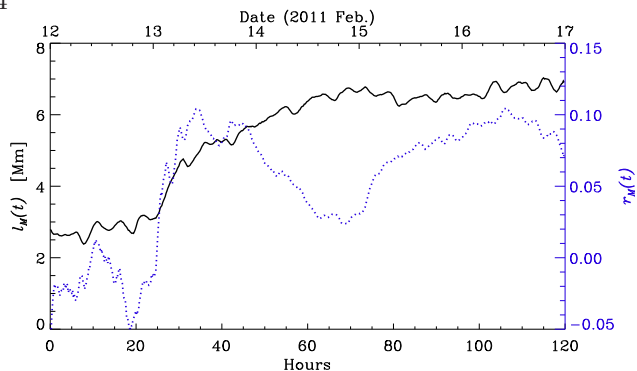}
\end{center}
\caption{
Evolution of $l_M$ (solid line) and $r_M$ (dotted line) for the
active region NOAA~11158.
}\label{fig:helispec5}
\end{figure}

The evolution of $l_M$ and $r_M$ in NOAA~11158 is shown
in Figure \ref{fig:helispec5}.
The average value of $r_M$ is about $+0.05$, while that of $l_M$ is about
$6\Mm$ in the developed stage of the active region.
These values are consistent with those quoted by \cite{Zhang2014}.
Being at $20^\circ$ southern heliographic latitude, the magnetic
helicity was found to obey the expected hemispheric sign rule (positive
in the south and negative in the north).
It is found that the fractional magnetic helicity $r_M$ shows a relatively
complex relationship with the temporal development of the active region and
it shows a similar tendency with ${\cal H}_M$ in Figure~\ref{fig:helispec3}.

\subsection{Active Region NOAA~11515}
 
\begin{figure}
\begin{center}
\includegraphics[width=0.5\textwidth]{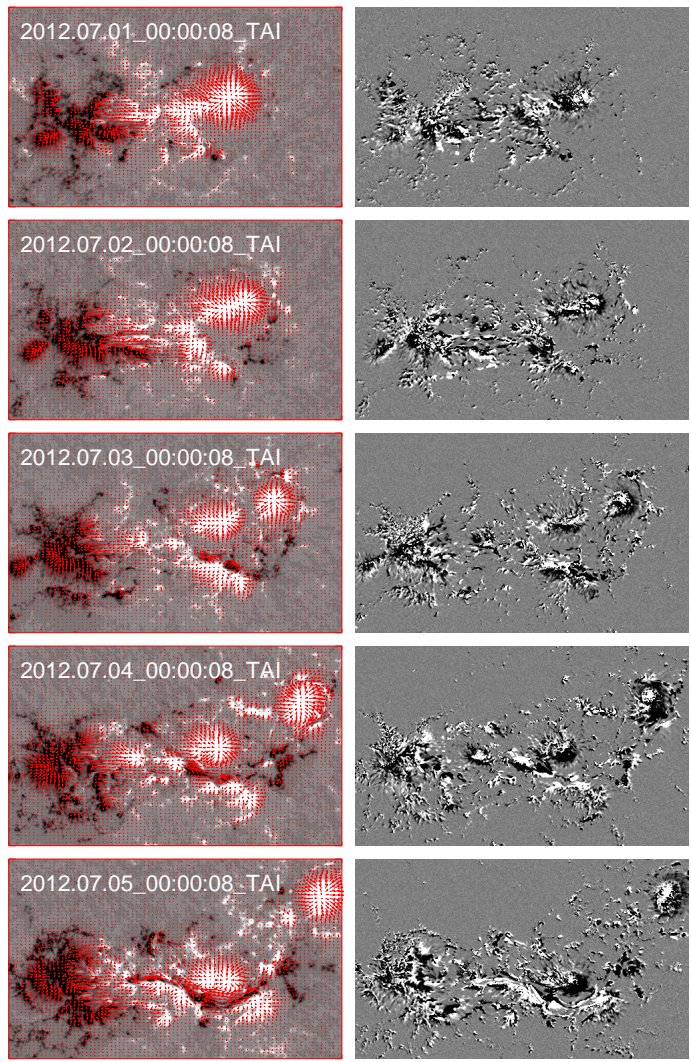}
\end{center}
\caption{
Photospheric vector magnetograms (left) and maps of $J_zB_z$ (right)
for the active region NOAA~11515 on 30 June -- 4 July 2012.
The arrows show the transverse component of the magnetic field.
Light (dark) shades indicate positive (negative) values of $B_z$ on the
left and $J_zB_z$ on the right.
}\label{fig:helispecs1}
\end{figure}

To study the evolution of magnetic energy and helicity spectra
in solar active regions, we have also analyzed HMI data
from the active region NOAA~11515 from 2012 June 30 to July 6
at approximately $18\degr$ southern latitude.
The pixel resolution of magnetograms is about $0.5''$, and the field of
view is $250''\times 150''$ in Figure~\ref{fig:helispecs1}.
In our study about 840 vector magnetograms have been used.

Figure~\ref{fig:helispecs1} shows the photospheric vector magnetograms and
the corresponding distribution of $h_C^{(z)}=J_z B_z$ from the vector
magnetograms of this active region on different days.
It shows the spatial distribution of the magnetic field and the current
helicity density of this active region at the solar surface.
 
\begin{figure}
\begin{center}\hspace*{-10mm}
\includegraphics[width=0.45\textwidth]{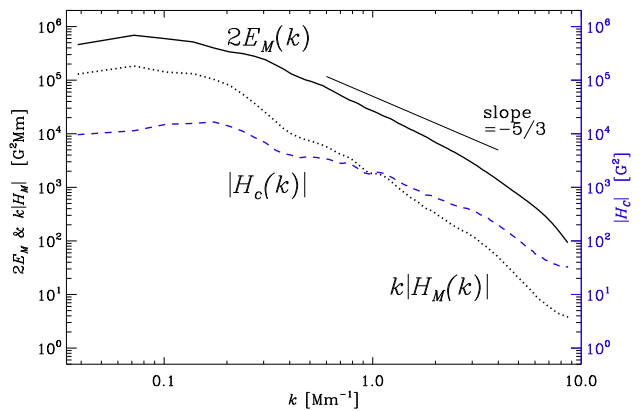}
\end{center}
\caption{
Mean spectra of $2E_M(k)$ (solid line), $k|H_M(k)|$ (dotted line),
and $|H_C(k)|$ (dashed line) obtained by averaging over 840 vector magnetograms of
active region NOAA~11515 on 2012 June 30 -- July 6.
}\label{fig:helispectots1}
\end{figure}

\begin{figure}
\begin{center}
\vspace*{4mm}
\hspace*{-2mm}
\includegraphics[width=0.48\textwidth]{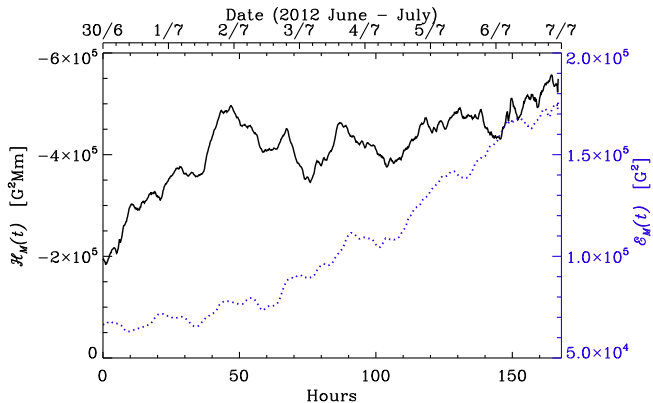}
\end{center}
\caption{
Evolution of mean photospheric magnetic helicity density
${\cal H}_M(t)$ (solid line) and mean magnetic energy density
${\cal E}_M(t)$ (dotted line) of active region NOAA~11515.
Note that the ordinate for ${\cal H}_M(t)$ shows negative values
increasing upwards.
}\label{fig:ene_helicd}
\end{figure}

To analyze some basic properties of the magnetic energy and helicity
spectra of this active region, we show in
Figure~\ref{fig:helispectots1} the averaged spectra that were inferred
from the vector magnetograms from 2012 June 30 to July 6.
These are comparable with the results of \cite{Zhang2014} and the average
spectrum of NAOO~11158 in Figure \ref{fig:helispectot1}.
Comparing the two active regions NOAA~11158 and 11515, we find that the
magnetic energy and helicity spectra are steeper in the latter case.

Figure~\ref{fig:ene_helicd} shows the evolution of mean magnetic
helicity and energy densities
of NOAA~11515, again obtained by integrating over all $k$.
First note that ${\cal H}_M(t)$ is negative even though NOAA~11515
is located at a southern heliographic latitude.
This is particularly surprising because $|{\cal H}_M(t)|$
is rather large, about 10 times larger than for NOAA~11158.
It is therefore unlikely that the surprising sign is a consequence
of fluctuations of a weak signal.
The mean magnetic energy density is about three to five times larger
than for NOAA~11158.
Similar to NOAA~11158, there is an intermediate phase (50--100 hr
after emergence) when ${\cal E}_M(t)$ still increases, but the increase of
$|{\cal H}_M(t)|$ is interrupted by a phase of varying
mean magnetic helicity density.

\begin{figure}
\begin{center}
\includegraphics[width=\columnwidth]{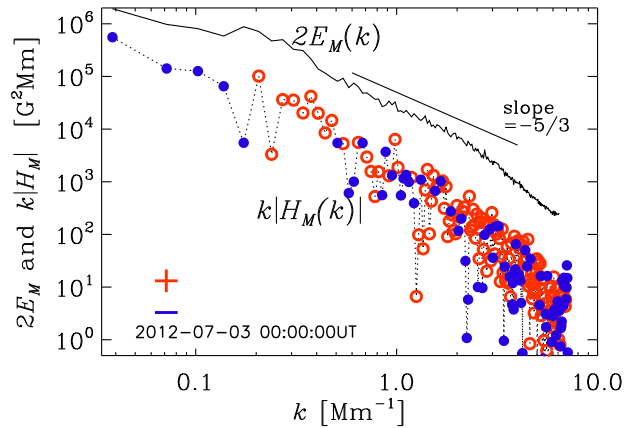}
\end{center}
\caption{$2E_M(k)$ (solid line) and $k|H_M(k)|$ (dotted line)
for NOAA~11515 at 0:00:00\,UT on 2012 July 3.
Positive (negative) values of $H_M(k)$ are indicated by open (closed)
symbols, respectively.
\label{fig:heli_phelicity_Bran_Zhq_11515}
}\end{figure}

To understand the origin of the unconventional sign of helicity in this
active region, we now consider the signed magnetic helicity spectra for
2012 July 3 where positive (negative) values are indicated by open (closed)
symbols.
It turns out that similar to NOAA~11158 where the signed magnetic
helicity spectrum was shown in Figure~2 of \cite{Zhang2014}, there is an
intermediate range of $0.2\Mm^{-1} \leq k\leq 0.5\Mm^{-1}$, where $H_M(k)$
consistently shows a positive sign just as expected for the small-scale
magnetic field in the southern hemisphere.
However, for $k < 0.2\Mm^{-1}$, the sign of $H_M(k)$ is in NOAA~11515
consistently negative (see Figure~\ref{fig:heli_phelicity_Bran_Zhq_11515}),
which agrees with the sign expected for the large-scale field;
see also \cite{BB03}.
Again, this is not so different from the case of NOAA~11158, which also
shows negative $H_M(k)$ for small $k$, but for NOAA~11515, the spectral
slope is much larger, which is the reason for the dominance of the
negative sign in the (integrated) mean magnetic helicity density.

By contrast, the current helicity, $\int k^2H_M(k,t)\,dk$, is
dominated by the high wavenumber end of the spectrum.
It is much noisier, so the hemispheric sign rule is often not obeyed.
Furthermore, as shown by \cite{Xu_etal15}, the isotropic approximation
usually fails in such a case.
While this must also be a concern for our present analysis, we should
emphasize that the systematic sign changes that are a function of $k$ are
certainly plausible and in agreement with theory \citep{BB03}.

\begin{figure}
\begin{center}
\vspace*{10mm}
\hspace*{-5mm}
\includegraphics[width=0.55\textwidth]{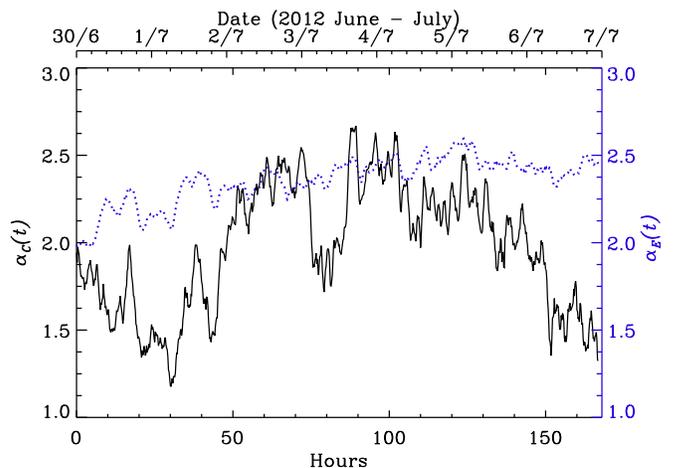}
\end{center}

\caption{Evolution of the spectral exponent $\alpha$ of the photospheric
current helicity spectrum (solid line) and the magnetic energy spectrum
(dotted line) of active region NOAA~11515.
}\label{fig:helispecs4}
\end{figure}

Figure~\ref{fig:helispecs4} shows the evolution of the mean spectral
indices $\alpha_E$ and $\alpha_C$ for NOAA~11515 using the same
wavenumber interval as before, i.e., $1\Mm^{-1}<k<6\Mm^{-1}$.
The minimum of $\alpha_C$ is 1.2, the maximum is 2.7,
and the mean value is about 2.0.
The minimum of $\alpha_E$ is 2.0, the maximum is 2.6,
and the mean value is about 2.4.
These values are larger than
those of NOAA~11158 and exceed the expected $5/3$ value.

\begin{figure}
\begin{center}
\vspace*{7mm}\hspace*{-6mm}
\includegraphics[width=0.55\textwidth]{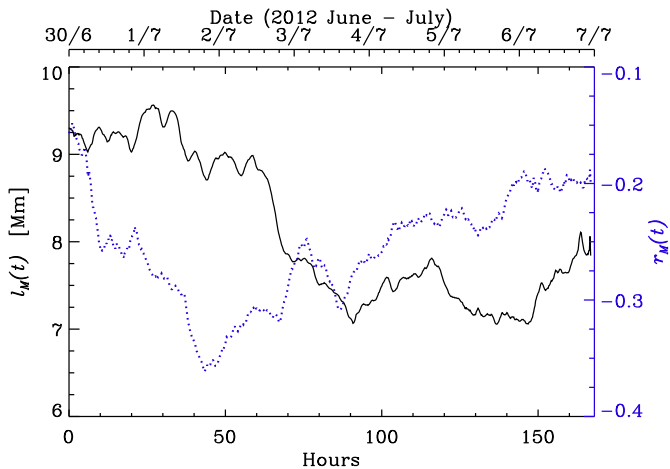}
\end{center}
\caption{
Evolution of $l_M$ (solid line) and $r_M$ (dotted line) for
active region NOAA~11515.
}\label{fig:helispecs5}
\end{figure}

The evolution of $r_M$ and $l_M$ is shown
in Figure~\ref{fig:helispecs5}.
The average value of $r_M$ is about $-0.26$ while that of $l_M$ is about
$8\Mm$ during the evolution of the active region.
Thus, the strength of $r_M$ is about five times larger than it is
for NOAA~11158.
We see that
the integral scale of the magnetic field decreases during the evolution
of the active region even though the area of the active region increases.
This is caused by a strong decrease of the mean magnetic energy density
and could be interpreted as a saturation mechanism whereby this active
region redistributes its rather large magnetic helicity over a larger area.

\section{Magnetic Helicity and Energy Spectra of Active Regions with the Solar Cycle}

\begin{figure}
\begin{center}\hspace*{-10mm} 
\includegraphics[width=0.44\textwidth]{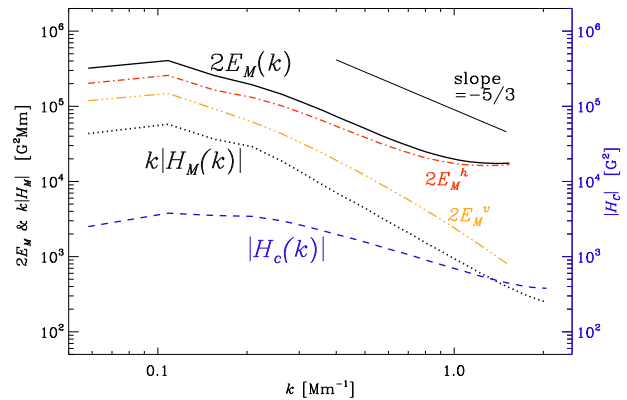}
\caption{
Averaged magnetic energy spectrum (solid line) together with
magnetic helicity (dotted line) and current helicity spectra (dashed line)
obtained from 6629 Huairou
vector magnetograms of solar active regions during 1988--2005.
$2E_M^{(h)}(k)$ (red, dash-dotted) and $2E_M^{(v)}(k)$
(orange, dash-triple-dotted) are shown for comparison.
}\label{fig:helispec6}
\end{center}
\end{figure}

Long-term statistical analyses of vector magnetograms at Huairou Solar
Observing Station have been obtained over recent years
\citep{Bao98,Gao08,zetal10}, covering the epochs of cycles 22 and 23.
These also provide an opportunity to analyze the evolution of the
variation of the spectra of magnetic fields of active regions and
the relationship with the solar activity cycle.
The averaged effect of active regions is important for the theoretical
interpretation and analysis of the solar cycle.
Figure~\ref{fig:helispec6} shows the averaged spectra of
$kH_M$ (dotted line), the current helicity $H_c$ (solid line),
and the magnetic energy $E_M$ (dashed line) using 6629 Huairou vector
magnetograms of the solar active regions observed from 1988 to 2005.
The method is equivalent to that for individual active regions above.
For consistency with the calculation of the long-term evolution
of the magnetic field of active regions by means of a series of Huairou
vector magnetograms, the spatial resolution of Huairou vector magnetograms
has been downsampled to $2''\times 2''$ to reduce the influence of the
different seeing conditions in the observations.
Due to the relatively low spatial resolution, the spectra shown in
Figure~\ref{fig:helispec6} become unreliable at large wavenumbers.
To substantiate this, we compare the estimated energy spectra based solely
on the horizontal fields as we did in \cite{Zhang2014}.
\begin{equation}
2E_M^{(h)}(k,t)=4\pi k\,\mbox{Re}\left\langle
\Gamma_{xx}+\Gamma_{yy}\right\rangle_{\phi_k},
\end{equation}
with those based solely on the vertical ones,
\begin{equation}
2E_M^{(v)}(k,t)=4\pi k\,\mbox{Re}\left\langle
\Gamma_{zz}\right\rangle_{\phi_k}.
\end{equation}
Note the factor $4\pi$ on the right-hand sides of the two
expressions above compared to only $2\pi$ in \Eq{EMdef}.
This accounts for why the two contributions should
give an estimate of the full spectrum $2E_M(k,t)$.
The shallow slope of the spectra of magnetic energy at high wavenumbers
is an artifact of the lower resolution and is found to mainly concern
the transverse components of the magnetic field.
While this may in fact be physical at least for the Huairou vector
magnetograms, the shallow spectral tails must be artifacts because they
are not reproduced with HMI at a higher resolution.
For more information about Huairou vector magnetograms we refer to the
papers by \cite{Ai86}, \cite{Su04a,Su04b}, and \cite{Gao08}.

\begin{table*}\caption{
Properties of active regions NOAA~11158 and 11515, and
average properties from Huairou Solar Observing Station.
}\centerline{\begin{tabular}{lccccrc}
\hline
\hline
 & Time & Latitude & $\alpha_E$ & $\alpha_C$ & $r_M$ & $l_M$ \\
\hline
NOAA~11158 & 2011 February 12 to 17   & $-13^\circ$ & $\quad$ 1.1 ... 1.7 ... 2.0 $\quad$ & 0.9 ... 1.6 ... 1.7 $\quad$ &  0.05 & $6\Mm$ \\
NOAA~11515 & 2012 June 30 to July 7 & $-18^\circ$ & $\quad$ 2.0 ... 2.4 ... 2.6 $\quad$ & 1.2 ... 2.0 ... 2.7 $\quad$ &$-0.26$& $8\Mm$ \\
Huairou    & 1980 to 2005           &  ---        & $\quad$         1.6         $\quad$ &         1.0         $\quad$ &       & $6$--$8\Mm$ \\
\hline
\hline\label{Tsummary}\end{tabular}}
\end{table*}

We find similar magnetic helicity and energy spectra as for
the individual active regions observed by HMI
and the averaged ones inferred from the active regions observed
at Huairou Solar Observing Station.
The variation of the mean current helicity of active regions inferred
from the Huairou vector magnetograms with sunspot butterfly diagrams has
been studied by \cite{zetal10}.
It shows the same tendency as the variation of helicity and energy
spectra for the individual active regions observed by HMI
and the averaged ones inferred from the active regions observed
at the Huairou Solar Observing Station; see Table~\ref{Tsummary}
for a more detailed comparison of the values quoted in Sections~3 and 4.

\begin{figure}
\begin{center}\hspace*{-14mm}
\includegraphics[width=0.41\textwidth]{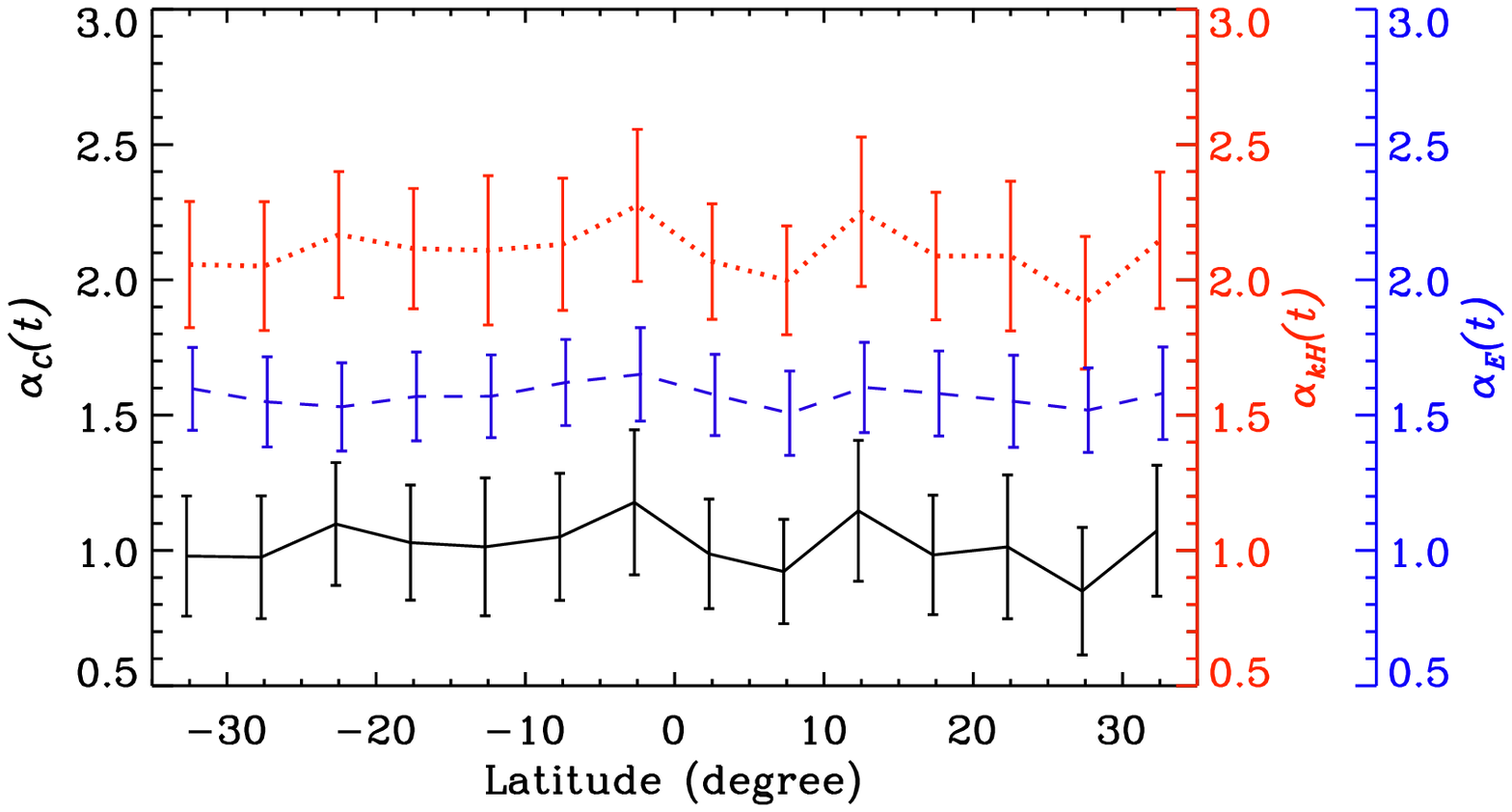}
\hspace*{-14mm}
\includegraphics[width=0.41\textwidth]{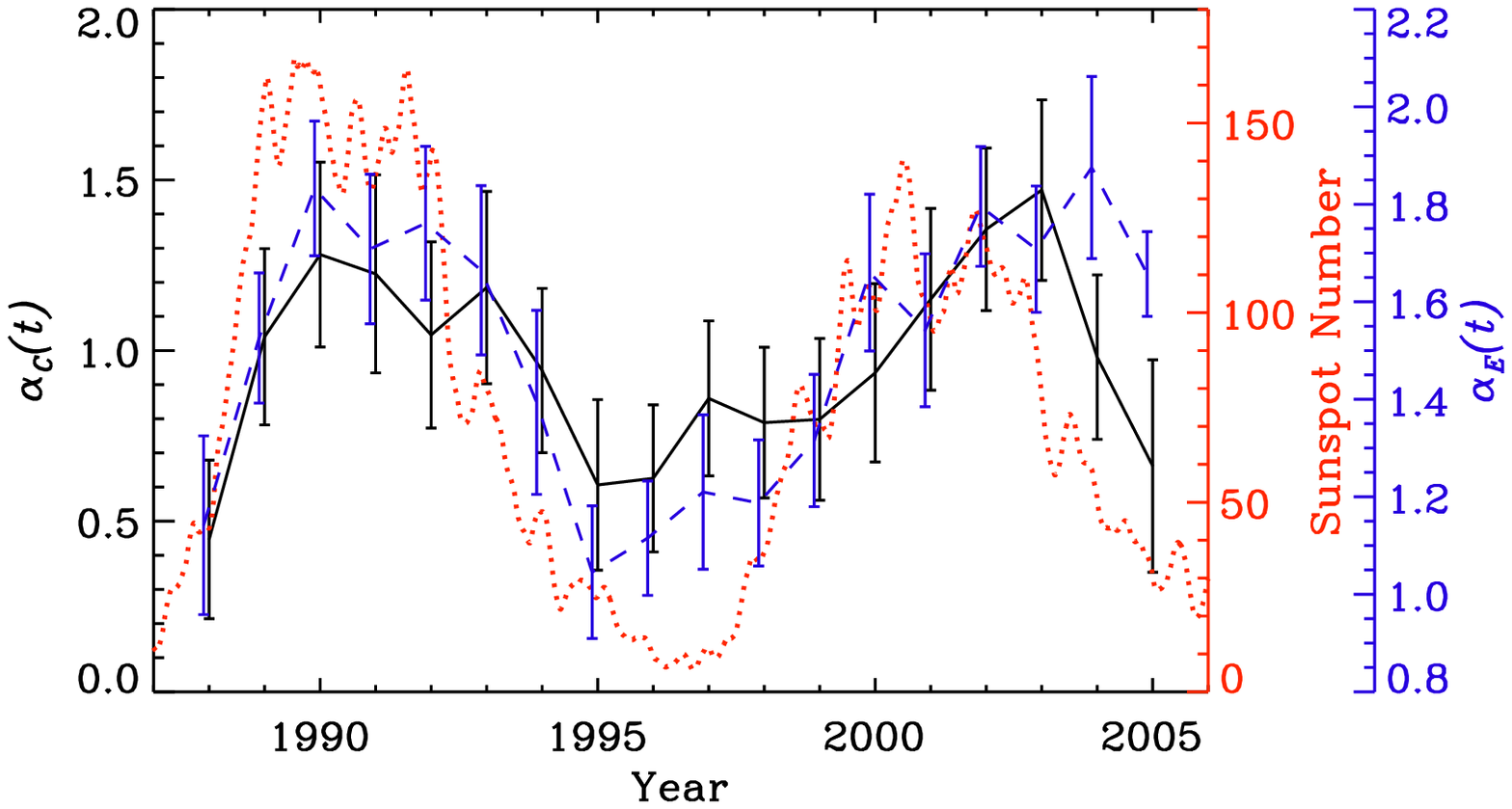}
\caption{
Top: variation of $\alpha_C$ (solid line) and $\alpha_E$ (dashed line)
with latitude (upper panel) and time (lower panel) inferred from 6629
Huairou vector magnetograms of solar active regions from 1988 to 2005.
The dotted line shows $\alpha_{kH}$ in the top panel and
sunspot number in the bottom panel.
The error bars are $0.3\sigma$.
}\label{fig:helispec7}
\end{center}
\end{figure}

To analyze the evolution of the averaged magnetic helicity and energy spectra
of solar active regions, Figure~\ref{fig:helispec7} shows the latitudinal
and temporal dependence of $\alpha_C$ and $\alpha_E$ with the solar cycle
in the spectral range of $0.2\Mm^{-1}<k<0.6\Mm^{-1}$.
The slopes of the spectra do not change systematically
with latitude when one averages the spectra of active regions for 1988 to 2005.
This suggests that the underlying mechanism for producing these fields
could be local small-scale dynamo action, which should operate equally
at all latitudes.
We emphasize that a small-scale dynamo is thought to operate independently
of the large-scale dynamo, i.e., equally well on all latitudes.
It is expected to produce scales that are smaller than those of the
energy-carrying eddies or the energy-carrying magnetic structures
$l_M$ \citep{BSS12}.
This scale is of the order of a megameter and thus much smaller than
the scale of active regions.
In reality both dynamos are coupled, and the small-scale dynamo could
even show a weak anticorrelation with the large-scale field \citep{KB15}.
Conversely, if we accept the local small-scale dynamo interpretation, the
constancy of the slopes would demonstrate the robustness of our method in
producing spectral slopes independent of seeing conditions and overall
magnetic field strength, which would be largest at low latitudes.
The mean spectral indices are 
$\alpha_{kH}\approx2.1$, $\alpha_C\approx1.0$, and $\alpha_E\approx1.6$.

Figure~\ref{fig:helispec7} shows the temporal variation of the
slopes of the spectra of magnetic energy and helicity of active
regions between 1988 and 2000.
These slopes show significant correlation with the sunspot number.
High values occurred from 1990 to 1992 and from 2000 to 2003, while
low values occurred during 1995.
These are consistent with the periods of solar maximum and minimum,
respectively.
The correlation coefficient between the slopes of the current
helicity spectra and sunspot numbers is 0.79 and that between the
magnetic energy spectrum and sunspot number is 0.77.
Note also that the mean magnetic energy density
during the solar maximum is high.
Furthermore, 
the maximum values of $\alpha_E$ and $\alpha_C$ tend to occur later
than the maxima of the sunspot number.
This is consistent with the observational result by \cite{zetal10} that
the maximum in the butterfly diagram of the mean current helicity
of active regions tends to be delayed compared with that of the sunspot number.
A similar indication is that the complex magnetic configuration
of active regions tends to occur in the decaying phase of solar cycle 23
(after 2002); see also \cite{Guo10}.

\begin{figure}
\begin{center}\hspace*{-20mm}
\includegraphics[width=0.45\textwidth]{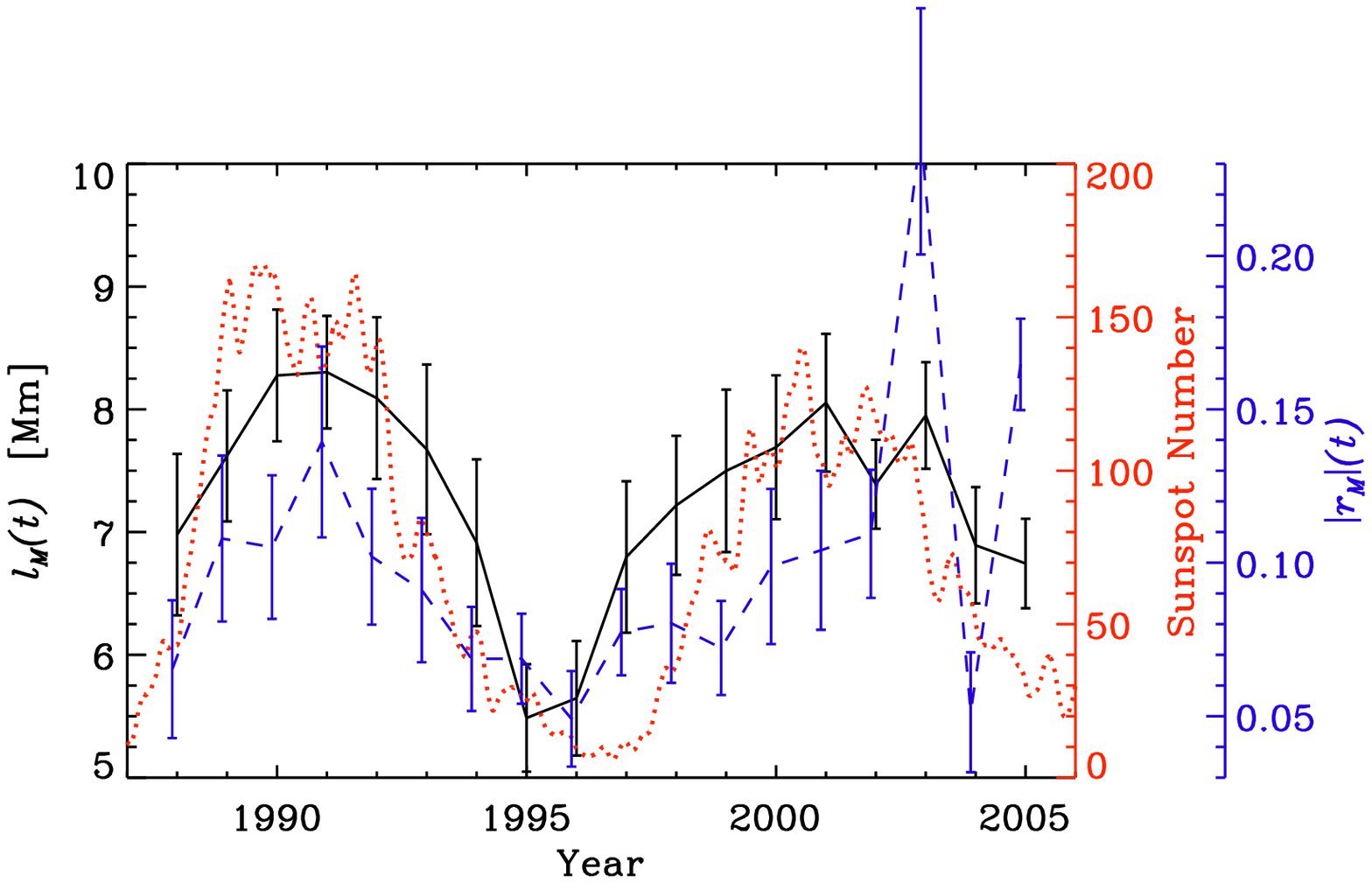}
\caption{
Solid line shows the temporal evolution of the integral scale $l_M$
of the magnetic field and the dashed line shows that of the photospheric
fractional magnetic helicity $|r_M|$ inferred by the 6629 Huairou vector
magnetograms of the solar active regions during 1988--2005.
The error bars are $0.3\sigma$.
The dotted line shows the sunspot numbers.
}\label{fig:helispec8}
\end{center}
\end{figure}

Figure~\ref{fig:helispec8} shows the temporal evolution
of the integral scale $l_M$ of the magnetic field of solar active
regions inferred from 6629 Huairou vector magnetograms during 1988--2005.
The correlation coefficient between the integral scale of the magnetic
field, inferred from \Eq{eq:integspec} and the sunspot numbers, is 0.80.
The average value of the integral scale of the magnetic field is about $8\Mm$
during solar maximum and $6\Mm$ during solar minimum for our calculated
active regions.
These dependencies are consistent with the finding that large-scale
magnetic patterns of active regions tend to occur near solar maximum.

Figure~\ref{fig:helispec8} shows that the averaged absolute values,
$\langle|r_M|\rangle$, of the photospheric fractional
magnetic helicity of active regions obtained from \Eq{eq:integspecr}
correlate with the solar cycle as measured by the sunspot number
except after 2003.
The peak in the mean relative magnetic helicity during 2003--2005 is
somewhat surprising, so we must ask about its physical significance.
In this connection it is interesting to recall that
based on analyses of MDI longitudinal magnetograms, thus using
different data sets, \cite{Guo10} reported an unusual magnetic field
distribution dominated by a few very strong active regions
during the declining phase of cycle 23.
In support of the physical significance of the peak, it should be emphasized
that there were several ``superactive'' regions such as NOAA~10484, 10486,
and 10488, especially near the end of 2003.
Of these, NOAA~10486 is generally associated with the famous Halloween
flare of 2003 October 28 \citep[e.g.][]{Hady09,Kazachenko10}.
However, all three of these active regions showed high
nonpotentiality \citep[cf.][]{liu06,Zhou07,Zhangy08a}. This is the reason
for the high fractional magnetic helicity $|r_M|$ occurring statistically
during this period.

\section{Resolution dependence}
\label{ResolutionDep}

In this study, the HMI and Huairou vector magnetograms have been used to
estimate the spectra of magnetic energy and helicity of solar active regions.
In addition, temporal changes of the magnetic energy spectra of active
regions and the evolution with the solar cycle have been found.
Since we use vector magnetograms of different spatial resolutions to
analyze the evolution of the spectral distributions of magnetic energy at
the different times, we now address the possible uncertainty regarding
the relationship between the observational resolution of the magnetic
field and the spectral shape at large wavenumbers.
The lower spatial resolution of vector magnetograms of
ground-based observations implies a source of error in the spectrum
of the magnetic field at high wavenumbers.

\begin{figure}
\begin{center}
\hspace*{-15mm}
\includegraphics[width=0.45\textwidth]{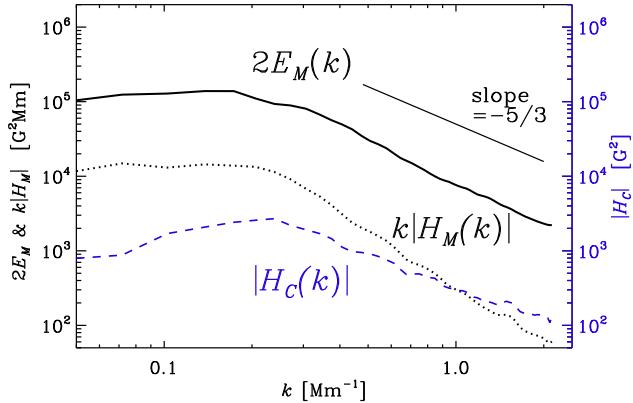}
\end{center}
\caption{
Same as Figure \ref{fig:helispectot1},
but with pixels compressed from 512$\times$512 to 128$\times$128.
}\label{fig:helispec13}
\end{figure}

\begin{figure}
\begin{center}
\hspace*{-15mm}
\includegraphics[width=0.45\textwidth]{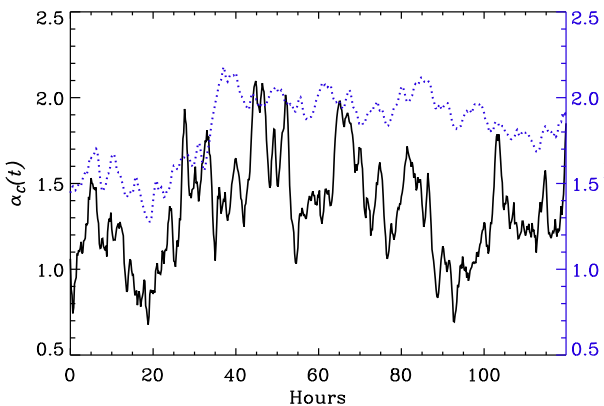}
\end{center}
\caption{
Same as Figure \ref{fig:helispec4},
but with pixels compressed from 512$\times$512 to 128$\times$128.
}\label{fig:helispec13b}
\end{figure}

To estimate the possible errors in the calculation of the magnetic spectrum
due to the low spatial resolution of the observational magnetic fields
by the Huairou vector magnetograms, Figure \ref{fig:helispec13} shows the
mean spectra of magnetic energy as well as magnetic and current helicity.
Figure~\ref{fig:helispec13b} shows
the evolution of the spectral indices $\alpha_C$ and $\alpha_E$
for wavenumbers in the active region NOAA~11158, whose pixel size
of the analyzed region of the HMI vector magnetograms have been downsampled
from $512\times512$ to $128\times128$.
The pixel resolution is $2''\times2''$, which is almost the same
as that of the Huairou vector magnetograms.
The same tendency is found for the magnetic energy spectra
as in Figure \ref{fig:helispectot1}.
The high noise in the timeseries of $\alpha_C$ and $\alpha_E$
in Figure \ref{fig:helispec4} is now reduced.
From 2011 February 12 to 16, the mean value of $\alpha_E$ is about 1.82 and
that of $\alpha_C$ is about 1.34 for $0.4\,{\rm Mm}^{-1}<k<2.0\,{\rm Mm}^{-1}$,
while the values obtained for the original resolution in
Figure~\ref{fig:helispec4} are 1.52 and 1.62, respectively
(the lower value of $\alpha_E$ given in Table~\ref{Tsummary}
is due to including the rapid growth during the emerging stage
of the active region on 2011 February 12).
For the detailed analysis we also have reversed the HMI vector
magnetograms to Stokes parameters ($Q$, $U$, and $V$) in the approximation
of the weak field with Gaussian smoothing for reducing to the forms of the lower
spatial resolution, compressing them to the lower pixel resolution of the Stokes
parameters, and then reverting to vector magnetograms again. We found almost the
same tendency for the spectrum of magnetic fields such as shown in Figure 15,
although the amplitudes of the slopes of the spectra of the magnetic fields
changed slightly. We notice that we still cannot imitate the real
case of the lower observational spatial resolution completely, such as the
shallow slope of the spectra of magnetic energy at high wavenumbers in
Figure~\ref{fig:helispec6}.
The difference with the degraded data implies that the resolution
of the observational vector magnetograms might
still be problematic in the diagnostics of the spectra of the magnetic field
in the detail study.
This may affect some of the analyses regarding the changes of the spectral
slopes with the solar cycle when using the Huairou vector magnetograms, although
one should keep in mind that our conclusions from the temporal variations
are compatible with those found for individual active regions.

\section{Conclusions}

We have applied the technique of \cite{Zhang2014} to estimate the magnetic
energy and helicity spectra using vector magnetogram data at the solar surface.
We have made use of the assumption that the spectral
two-point correlation tensor of the magnetic field can be approximated by
its isotropic representation.
In this paper we have analyzed the evolution of magnetic energy and
helicity spectra in active regions and have also analyzed the changes
during the solar cycle.
Our major results are the following.

\begin{enumerate}
\item
The values of $\alpha_E$ and $\alpha_C$ of solar active regions
are of the order of $5/3$, although $\alpha_C$ is slightly smaller than
$\alpha_E$, i.e., the current helicity spectrum is slightly shallower
than the magnetic energy spectrum.
We have also found a systematic change of $\alpha_E$ and $\alpha_C$ with the
temporal development of active regions, which reflects their structural changes.

\item
There is not necessarily an obvious relationship between the change of the
photospheric fractional magnetic helicity $r_M$ and the integral scale
of the magnetic field $l_M$ of individual active regions.
Nevertheless, Figures~\ref{fig:helispec5} and \ref{fig:helispecs5}
show that $r_M$ and $l_M$ tend to be correlated most of the time, which is
also true for the cyclic variation shown in Figure~\ref{fig:helispec8}.
Looking at Equation~\ref{rM_and_lM}, this might be somewhat surprising
because it shows that $r_M$ and $l_M$ should be anticorrelated if
${\cal H}_M$ and ${\cal E}_M$ were constant.
However, neither of them are constant and both show significant
variations (see Figures~\ref{fig:helispec3} and \ref{fig:ene_helicd}).

\item
We have found that there is a statistical correlation between the variation
of the spectra of magnetic energy and helicity of solar active regions
with the solar cycle.
This implies that there is a trend for the characteristic scales and
the intensity of the magnetic field of the active regions to increase
statistically with the solar cycle.

\item
Interestingly, even through the mean $\alpha_E$ and $\alpha_C$ of active
regions vary with the cycle and increase with increasing mean
magnetic energy density, they do not change with latitude even though
the mean magnetic energy density does change with latitude.

This suggests that the underlying magnetic field represents a part that
is independent of the global cyclic magnetic field and possibly a
signature of what is often referred to as local small-scale dynamo.

\item
In NOAA~11515, where the fractional magnetic helicity is rather
large (with a peak value of 35\% and 26\% on average), the integrated
mean magnetic helicity density has the opposite sign of what is
expected for its hemisphere.
This is associated with a steepening of the magnetic helicity spectrum
at large scales, therefore giving a preference to the contributions of the
large-scale field whose magnetic helicity is indeed expected to have
the opposite sign.

\end{enumerate}

Values of $\alpha_E$ and $\alpha_C$ of the order of $5/3$ are roughly
compatible with a Kolmogorov-like forward cascade \citep{Ko41,Obukhov41},
which is expected from the theory of nonhelical hydromagnetic turbulence
when the magnetic field is moderately strong \citep{GS95}.
However, for decaying turbulence \cite{Lee2010} found that the scaling
depends on the field strength and takes on a shallower Iroshnikov--Kraichnan
$k^{-3/2}$ spectrum \citep{Iro63,Kraichnan65} for weaker fields and a
steeper $k^{-2}$ weak-turbulence spectrum for stronger fields;
see \cite{BN11} for the respective phenomenologies in the three cases.
The steeper $k^{-2}$ spectrum has also recently been found in decaying
turbulence simulations where the flow is driven entirely by the magnetic
field \citep{BKT15}.
It is thus tempting to associate the changes in the values of $\alpha_E$ and
$\alpha_C$ with corresponding changes between these different scaling laws.

\acknowledgments

We thank the referee for constructive criticism that has led to
many improvements in the presentation.
We gratefully acknowledge the {\em SDO} team for the HMI vector magnetograms
and Drs.\ Yu Gao, Haiqing Xu, and K.\ Kuzanyan for the data processing
of the Huairou vector magnetograms.
We also thank Dr.\ Dhrubaditya Mitra for suggesting a connection with
different hydromagnetic scaling laws for different magnetic field strengths.
This study is supported by grants from the National Natural Science
Foundation (NNSF) of China under the project grants 10921303, 11221063,
and 41174153 (HZ), the NNSF of China and the Russian Foundation for
Basic Research under the collaborative China-Russian projects 13-02-91158 and 15-52-53125
(HZ+DDS), and the Swedish Research Council under the project grants
2012-5797 and 621-2011-5076 (AB).

\newcommand{\pmn}[2]{ #1, {MNRAS}, in press, arXiv:#2}
\newcommand{\papj}[2]{ #1, {ApJ}, in press, arXiv:#2}
\newcommand{\yapj}[3]{ #1, {ApJ,} {#2}, #3}
\newcommand{\yjfm}[3]{ #1, {JFM,} {#2}, #3}
\newcommand{\ymn}[3]{ #1, {MNRAS,} {#2}, #3}
\newcommand{\yana}[3]{ #1, {A\&A,} {#2}, #3}
\newcommand{\yapjl}[3]{ #1, {ApJL,} {#2}, #3}
\newcommand{\yprl}[3]{ #1, {PhRvL,} {#2}, #3}
\newcommand{\yprd}[3]{ #1, {PhRvD,} {#2}, #3}
\newcommand{\ypre}[3]{ #1, {PhRvE,} {#2}, #3}
\newcommand{\yjour}[4]{ #1, {#2,} {#3}, #4}
\newcommand{\yjswsc}[3]{ #1, {JSWJC,} {#2}, #3}
\newcommand{\ybook}[3]{ #1, {#2} (#3)}

\end{document}